\documentclass[a4paper]{article}

\usepackage[USenglish]{babel} 
\usepackage[T1]{fontenc}
\usepackage[ansinew]{inputenc}

\usepackage{lmodern} 

\usepackage{graphicx} 

\usepackage{amsmath}
\usepackage{amsthm}
\usepackage{amsfonts}

\begin{document}

\pagestyle{empty} 


\title{Coulomb gauge QCD from flow gauge at three loops}

\author{A Andra\v si$^+$ and J C Taylor\footnote{Corresponding author } 
\footnote{\textit{E-mail addresses} aandrasi@irb.hr (A. Andrasi), jct@damtp.cam.ac.uk (J. C. Taylor)} \\ \\ {\it  $^+$Vla\v ska 58, Zagreb, Croatia} \\ $^\dagger${\it DAMTP, Cambridge University, UK}}





\maketitle



\begin{abstract}

\noindent{We study the consistency of the non-Abelian Coulomb gauge. There are
energy divergences in individual diagrams, which are known to cancel to 2-loop order when suitable sets of graphs are summed. We investigate  
to 3-loop order the inclusion of UV divergent sub-graphs into the energy divergences. In all the examples we study, we find sets of graphs which are free of energy divergences. We make use of an interpolating gauge to regularize the energy divergences while integrals are manipulated. We comment on radiative corrections to the Christ-Lee terms in the Hamiltonian.}\\

\noindent{Pacs numbers: 11.15.Bt; 03.70.+k}\\

\noindent{Keywords QCD, Coulomb gauge}

\end{abstract}

\vfill\newpage

\pagestyle{plain} 

\section{Introduction}
The Coulomb gauge in non-abelian gauge theory is the only explicitly unitary gauge.
But in perturbation theory it suffers from 'energy divergences', that is Feynman integrals which are divergent  over the
time-components of the momenta, while the spacial components are held fixed.
The simplest such energy divergences occur at one loop. For pure YM theory, these are quite easily canceled by combining Feynman
diagrams appropriately, but they are automatically removed by using the Hamiltonian, rather than the Lagrangian formalism \cite{mohapatra}.
When quark loops are included, Ward identities secure the cancellation of this type of energy-divergence \cite{aajct1}.

More subtle divergences appear first at two loop order. The cancellation of these was proved by Doust\cite{doust} (see also \cite{cheng}
and generalized in \cite{aajct2}).
The origin of these divergences was linked by Christ and Lee \cite{christlee} with the problem of correctly ordering the factors in the
Coulomb potential in the Hamiltonian (see also \cite{schwinger}). But in this paper we consider only ordinary momentum-space Feynman perturbation theory,
in the manner of Doust.

 Pure energy-divergences, that is divergences over the energy integrals with all spatial momenta fixed, occur
at 2-loop order only, not at higher order (see \cite{doust}). But if ordinary UV divergences are combined with energy divergences, new problems occur at 3-loop order. These are the subject of this paper. Specifically we study the insertion of  UV divergent quark loops into
two-loop gluon graphs. Can the divergences still be canceled by judiciously combining Feynman graphs?
A difficulty in attempting this is to be sure that the divergent integrals we are manipulating are well-defined.
To overcome this problem, we make use of a 'flow gauge', which interpolates between the Feynman gauge
and the Coulomb gauge. This flow gauge is characterized by a parameter $\theta$, $\theta=1$ is the Feynman
gauge, and the  Coulomb gauge is defined by the limit $\theta \rightarrow 0$. For nonzero $\theta$, there are no energy divergences
in any Feynman integral. We want to show that, for suitable combinations of graphs, the limit $\theta \rightarrow 0$ yields convergent
integrals. We emphasize that the flow gauge is of no practical use, having  the advantages of neither the Feynman nor the Coulomb gauge.
We use it only as a mathematical tool.

In the 3-loop graphs we consider, there are energy divergences in individual graphs (the UV divergences in sub-graphs having been removed by renormalization). We study six examples. In each case we are able to identify sets of graphs such that in their sum the integrand is well behaved at high energies in the limit $\theta \rightarrow 0$. Unfortunately, we have not been able to prove a general theorem in the manner of Doust \cite{doust}. For all but one of our examples, we find a fairly simple closed form for the sum of the set of energy-divergent graphs.

To 2-loop order, Doust has shown that the summed energy divergences give the $O(\hbar^2)$ terms in the Hamiltonian which were derived by Christ and Lee by consideration of operator ordering.  The 3-loop energy divergences  give higher order corrections but they are not energy-independent like the  Christ-Lee operator.
We are able to identify  simple contributions to these corrections.

\section{The interpolating gauge}
We use indices $i,j,k,l,m,n$ for spatial vectors; $\lambda, \mu, \nu$ for Lorentz vectors, $a,b,c,d$ for colour.

Energy divergences occur when there are integrals which are divergent over the energy variables, with the spatial momenta held fixed.
These divergences are removed by going to a gauge defined by the gauge-fixing term
\begin{equation}
-\frac{1}{2\theta^2}[(\partial_i A^i+\theta^2 \partial_0 A^0)^2].
\end{equation}
 For $\theta=1$ this gauge is the Feynman gauge, and the limit
$\theta \rightarrow 0$ gives the Coulomb gauge. We use this gauge only as a mathematical tool, so that we are dealing with well-defined integrals in the progress of the work.

 We will use  the notation that the momentum $k=(k_0,\textbf{K})$ and $k^2=k_0^2-\textbf{K}^2$, and
\def\K{\bar{K}}
\begin{equation}
\K^2\equiv \textbf{K}^2-\theta^2 k_0^2.
\end{equation}
\def\c{\gamma}
\def\t{\theta}
\def\R{\bar{R}}
\def\P{\bar{P}}
\def\Q{\bar{Q}}
In the gauge given by (1), the Coulomb propagator and the ghost propagator are both 
\begin{equation}
-\frac{1}{\K^2},
\end{equation}
the spatial propagator is
\begin{equation}
\frac{1}{k^2+i\eta}\left[g_{ij}+\frac{K_i K_j}{\K^2}\right].
\end{equation}
Since we use the Hamiltonian formalism, we require also propagators involving the electric field $\bf{E}$. It is
\begin{equation}
\frac{K^2}{k^2+i\eta}\left[ g^{mn} +\frac{K^m K^n}{\textbf{K}^2} \right].
\end{equation}
(unlike (4), this is transverse.) (We use indices $i,j,...$ for the potential $\bf{A}$ and indices $m,n,...$ for $\bf{E}$).
There are also off-diagonal propagators. That between $E^m$ and $A_j$ is
\begin{equation}
\frac{ik_0}{k^2+i\eta}\left[ g^m_j+\frac{K^mK_j}{\K^2} \right].
\end{equation}
and that between $E_m$ and $A_0$ is
\begin{equation}
\frac{iK^m}{\K^2}.
\end{equation}
\begin{figure}[t]
		\includegraphics[width=0.5\textwidth]{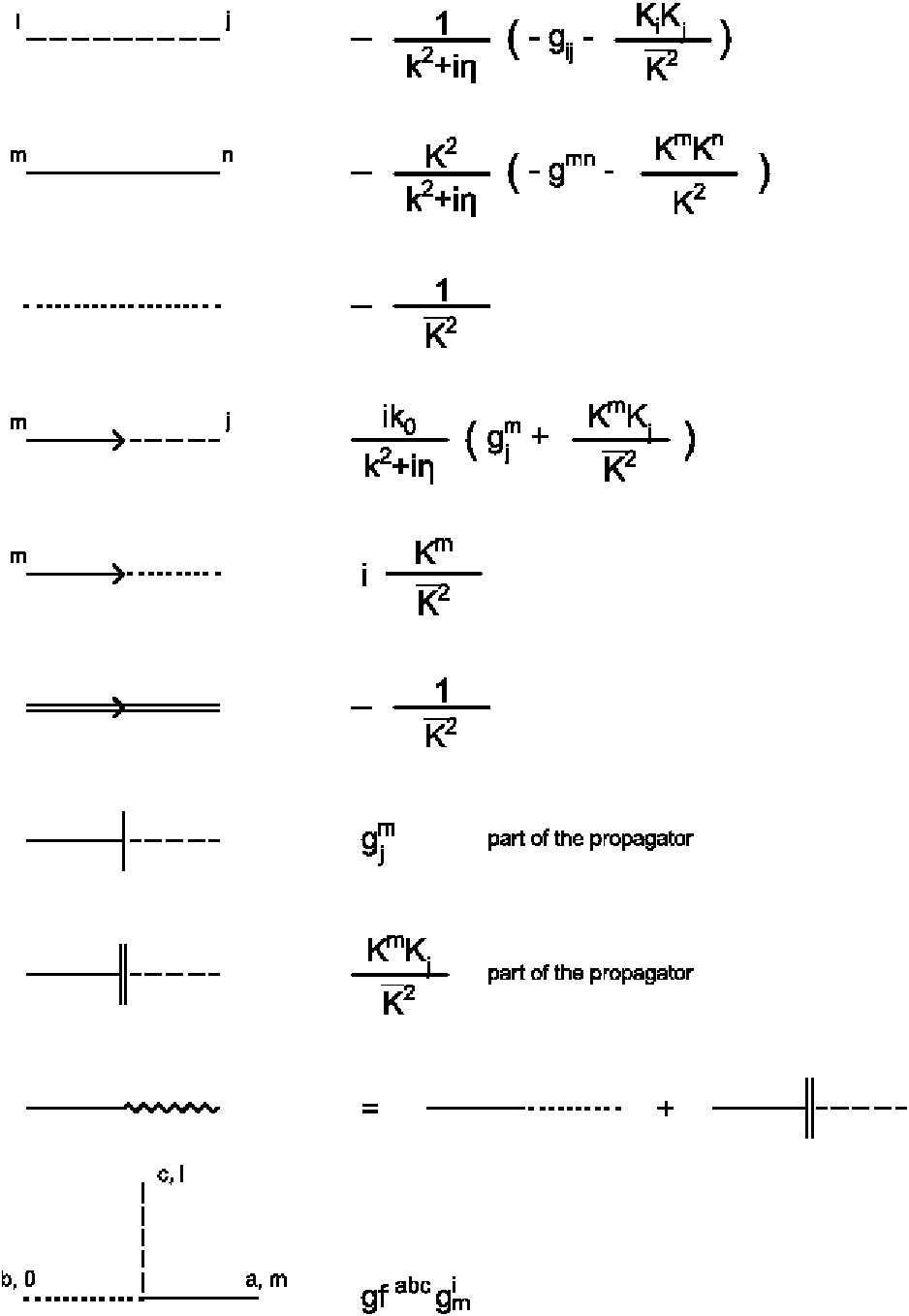}
	\label{fig:Fig1a.eps}
\hspace{20 pt}
		\includegraphics[width=0.5\textwidth]{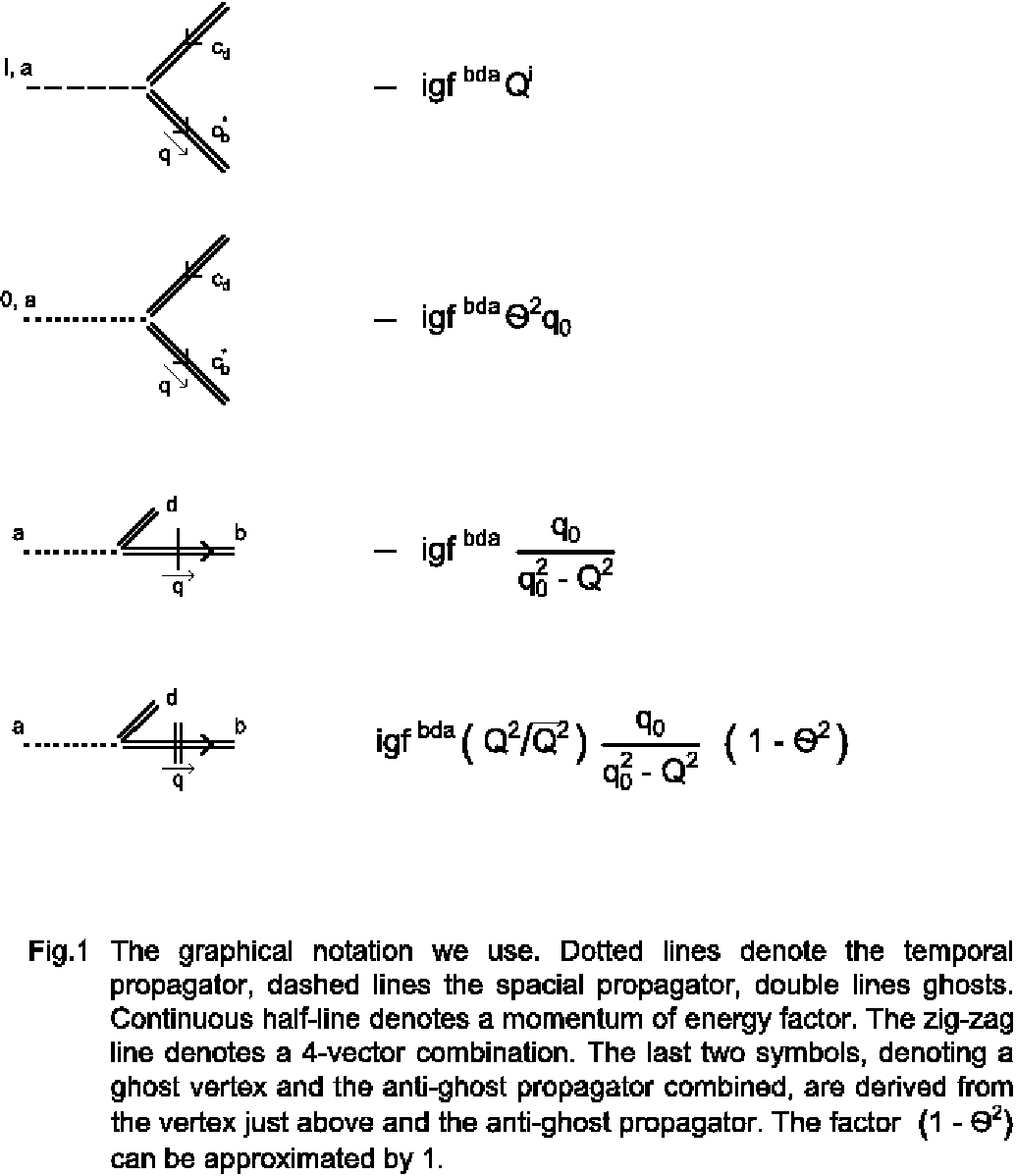}
	\label{fig:Fig1b}
\end{figure}
Our graphical conventions for these propagators are shown in Fig1a and Fig1b. We also use a graphical notation for parts of the propagators
which have different numbers of $\t$-dependent denominators, and also a notation for a combination which is approximately proportional to a 4-vector like
$p_{\mu}$, since this enables us to make use of  Ward identities for the quark loops (see the Appendix). 

In the limit $\theta \rightarrow 0$ the above Feynman rules reduce to the Feynman rules of  the Coulomb gauge (see for example \cite{aajct4}). We will take this limit only after we have identified groups of graphs which
are free of energy divergences in the limit.
\section{Energy divergences at one and two loop order}
At one loop order, there are  terms which are linearly divergent when $\t=0$ in the Lagrangian formalism. An example is 
\begin{equation}
\int dp_0\frac{(2p_0k_0)^2}{(k-p)^2 \P^2},
\end{equation}
These are not present in the Hamiltonian formalism, which we  use.

To two loop order there are logarithmic energy divergences (in the limit $\t=0$) in forms like
\begin{equation}
\int dp_0 dq_0  \frac{p_0q_0}{(p^2+i\eta)(q^2+i\eta)}f(\P^2,\P'^2,\Q^2,\Q'^2,\R^2).
\end{equation}
 Doust has proved \cite{doust} that graphs can be combined so that these
appear in the combination
\begin{equation}
\int dp_0 dq_0 dr_0 \delta(p_0+q_0+r_0-k_0)Df(\P^2,\P'^2,\Q^2,\Q'^2,\R^2)
\end{equation}
where
\begin{equation}
D\equiv \left[\frac{p_0q_0}{p^2q^2}+\frac{q_0r_0}{q^2r^2}+\frac{r_0p_0}{r^2p^2}\right]
\end{equation}
(we omit the Feynman $i\eta$ for shortness),  and since the integral is convergent we can take the limit as $\t \rightarrow 0$ and get (see equations (4.1) and (4.2) of \cite{doust}) 
\def\p{\textbf{P}}
\def\q{\textbf{Q}}
\def\r{\textbf{R}}
\def\k{\textbf{K}}
\begin{equation}
-\pi^2 f(\p^2,\q^2,\p'^2,\q'^2,\r^2),
\end{equation}
independent of energies.
\def\U{\frac{p_0q_0}{p^2q^2}}
\def\V{\frac{r_0p_0}{r^2p^2}}
\def\W{\frac{q_0r_0}{q^2r^2}}

\def\u'{\frac{p'_0q'_0}{{p'}^2{q'}^2}}
\def\v'{\frac{r_0p'_0}{r^2{p'}^2}}
\def\w'{\frac{q'_0r_0}{{q'}^2r^2}}

An important result which we will use is that single integrals like
\begin{equation}
\int dp_0 \frac{p_0}{p^2+i\eta}\frac{1}{\P^2}=0,
\end{equation}
where the second factor makes the integral convergent for $\t \neq 0$.

\section{A simple 2-loop example} 
Although the 2-loop energy divergences are well understood, see  \cite{doust}, we exhibit the simplest example, to make some points clear, and for comparison in later sections.
\begin{figure}[th]
	\centering
		\includegraphics[width=0.75\textwidth]{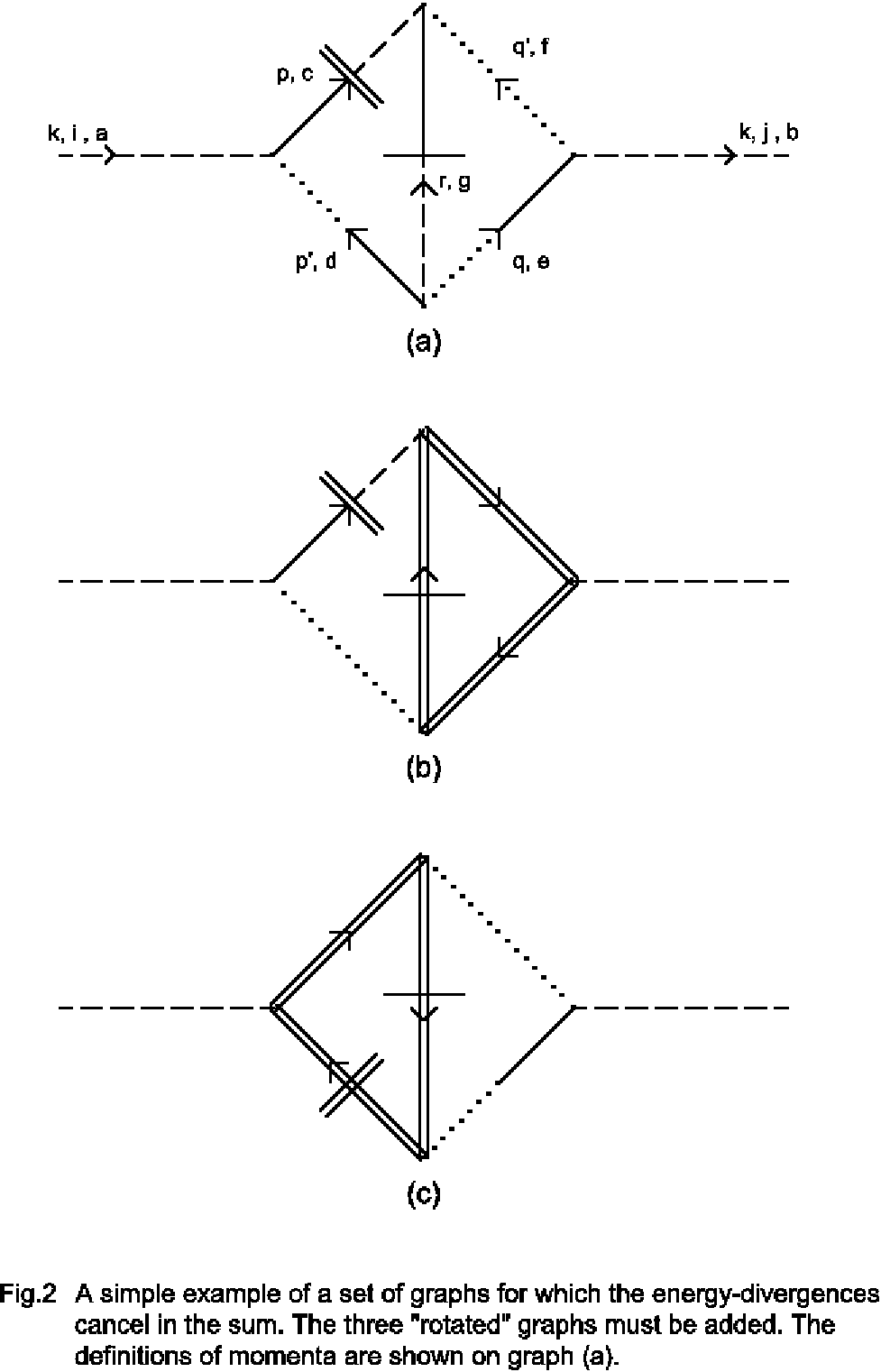}
	\label{fig:Fig2}
\end{figure}

Our example is the graph in Fig.2(a) which gives the integral
\begin{equation}
-\c\int d\textbf{P} d\textbf{R} \int dp_0 dq_0 \V \textbf{P}.\textbf{P}' [\P^2\Q^2\P'^2\Q'^2]^{-1}P_iQ_j,
\end{equation}
where $p+q+r=k$ and
\begin{equation}
\c=\frac{1}{2}g^4(2\pi)^{-4}C_G^2 \delta_{ab}
\end{equation}
$a,b$ being colour indices and $C_G$ the colour group Casimir.

If we take $\t=0$, the $q_0$ sub-integration (with $p_0$ fixed)  is logarithmically divergent, and the complete integral is not well defined.
With $\t \neq 0$, however, the result of the $q_0$ sub-integration ($p_0$ fixed) is something proportional to $\t $. This may be seen by  changing of the variable $\hat{q}_0=\t q_0$ when the only $\t$-dependence in the sub-integrand is in the factor
\begin{equation}
\frac{\t p_0+\hat{q}_0-\t k_0}{(\t p_0-\t k_0+\hat{q}_0)^2-\r^2},
\end{equation}
which is an odd function of $\hat{q}_0$ when $\t p_0 =0$. Thus  the one-loop graph is zero in the Coulomb gauge (defined as the limit of the flow gauge). But, for the 2-loop graph,  if the $p_0$-integration is done as well (before letting $\t \rightarrow 0$), there are contributions from $p_0=O(1/\t)$ giving a non-vanishing final result. This
 result looks nothing like the Coulomb gauge.

 The usual nesting property of Feynman integrals, that sub-graphs are graphs in their own right, breaks down.

The cure for this trouble is to combine suitable sets of graphs before taking the limit $\t \rightarrow 0$.  In the present example,
we must add the other graphs Fig.2(b),(c)  (containing ghost loops, as it happens) which replaces $\V{\p.\p'}$ in (14) by
\begin{equation}
 {\V} \p.\p'+{\V} \p.\q'-{\v'} \p'^2.
\end{equation}
Approximating $\v'$ by $\V$ (a step which we discuss below), (17) gives
\begin{equation}
 \V X_0 \equiv \V \frac{1}{2}[\r^2-\k^2-\p'^2-\q'^2].
\end{equation}
We add the rotated graphs, obtained by rotating the internal lines in the plane of the paper through 180 degrees, which amounts to the substitutions
\begin{equation}
p, q \rightarrow q, p.
\end{equation}
Then, since (18) is symmetric under (19), the result  contains a factor
\begin{equation}
\V +\W,
\end{equation}
and using (13) we can add on $\U$, thus getting, in terms of  the convergence factor $D$ defined in (11),
\begin{equation}
\c\int d\textbf{P}d\textbf{Q}dp_0dq_0DX_0[\P^2\Q^2\P'^2\Q'^2]^{-1}.
\end{equation}
 Then the limit $\t \rightarrow 0$ can be taken, which means replacing $\P$ by $\p$, etc. Finally. the energy integrals may be done giving
\begin{equation}
-\pi^2 \c\int d\textbf{P}d\textbf{Q}X_0[\p^2\q^2\p'^2\q'^2]^{-1}.
\end{equation}
The result (22) can be viewed in either of two ways. It can be viewed as being derived from Coulomb gauge Feynman integrals, as above,
by judiciously combining sets of graphs, in the manner of \cite{doust}. Or one can insert the result as a new $O(\hbar^2)$ operator, called $V_1 +V_2$,   into the Hamiltonian, as Christ and Lee did \cite{christlee}, and at the same time make a prescription that potentially divergent Feynman 2-loop integrals like (14) are to be set zero. The latter would probably be the simpler approach in practice. (With dimensional regularization, there are no UV divergences in $V_1+V_2$, because it is like 3-dimensional field theory and the only poles are at even values of $d-1$).

We must now justify the approximation used above of replacing $\v'$ by $\V$ in (17). The neglected piece involves the integral
\begin{equation}
\int dr_0  \frac{r_0}{r_0^2-\r^2}\int dp_0\left[\frac{p'_0}{p_0^{'2}-\p'^2}-\frac{p_0}{p_0^2-\p^2}\right][\P^2\P'^2\Q^2\Q'^2]^{-1}.
\end{equation}
If the $p_0$ integration is done (for fixed $r_0$) the result is a function of $\t^2 r_0^2, \t^2 r_0 k_0, \t^2 k_0^2$ and of the spatial vectors. Since $k_0$ is a fixed quantity, we can let $\t k_0 \rightarrow 0$, but since $r_0$ ranges from $-\infty$ to $+\infty$ we cannot neglect $\t r_0$. Then the $r_0$-integral in (23) is convergent and, since the integrand is odd, is zero. Thus the correction is zero in the limit $\t \rightarrow 0$. (This argument was not made explicitly in \cite{doust} but was made in \cite{aajct2}.)
We will use this argument several times below. It applies whenever there is a term like $\V$, and other factors render either the
$p_0$-  or the $r_0$-integral convergent even with $\t=0$.

\section{Energy divergences in some three-loop graphs}
The purpose of this paper is to investigate energy divergences at three loop order. We examine only the particular case where UV-divergent quark loops are inserted into two-loop gluon graphs. This case is simplified because the quark loops obey simple Ward identities rather than the more complicated BRST identities. We use the Ward identities to relate the high energy behaviour of the different quark loops. In fact
the high energy behaviour of all quark loops can be expressed in terms of the gluon self-energy quark loop, $S(p)$, as was shown in \cite{aajct3}.
The results of \cite{aajct3} are summarized in the Appendix.
We show that graphs can be grouped into sets, for each of which we get integrals of the form of (11), and also contain the self-energy quark loop functions $S(p^2)$, $S(r^2)$, etc (defined in the Appendix). These integrals are then convergent at high $p_0, q_0, r_0$ when $\t=0$. 
But in general, since we know only the high energy limit, that is 
\begin{equation}
p_0,q_0,r_0 >> |\textbf{P}|,|\textbf{Q}|,|\textbf{R}|,|\textbf{K}|,k_0.
\end{equation}
   we can show only that the form (10) is correct in the high energy range of the integration. There are  corrections suppressed at high energies, which therefore have  no energy divergence. Some of these corrections can be shown to be zero, by the argument used at the end of section 4. But others are non-zero, but can be  evaluated fairly simply.
Leaving aside these corrections, the final step is to combine graphs to give an energy-integral of the form
\begin{equation}
\c\int dp_0 dq_0 dr_0 \delta(p_0+q_0+r_0-k_0)\left[\frac{p_0q_0}{p^2q^2}+\frac{q_0r_0}{q^2r^2}+\frac{r_0p_0}{r^2p^2}\right]S(r^2)f(\p^2,\p'^2,\q^2,\q'^2,\r^2).
\end{equation}
(Or the same form but containing $S(p')$.)
The limit $\t \rightarrow 0$ still exists, but the energy integrals in (25) are no longer trivial, and the result is not independent of $k_0$,
as is the Christ-Lee operator.

All the graphs which we study  give an integral of the form
\begin{equation}
\c\int d\textbf{P}d\textbf{Q}dp_0dq_0F_{ij}(\textbf{P},\textbf{Q},\textbf{R},\textbf{P}',\textbf{Q}';p_0,q_0,r_0,p'_0,q'_0)
\end{equation}
where $\textbf{P}'=\textbf{P}-\textbf{K}$, $\textbf{Q}'=\textbf{Q}-\textbf{K}$, $p'_0=p_0-k_0$, $q'_0=q_0-k_0$.

There are two changes of variables which can be made when they are useful:
\begin{equation}
p,q,r \leftrightarrow -p',-q',-r,
\end{equation}
and
\begin{equation}
p, q, r  \leftrightarrow q, p, r.
\end{equation}

There are six independent scalars which can be constructed from the three 3-vectors $\textbf{P,Q,K}$. We choose the set
\begin{equation}
\textbf{K}^2, \textbf{R}^2, \textbf{P}^2, \textbf{P}'^2, \textbf{Q}^2, \textbf{Q}'^2
\end{equation}
The numerators of the Feynman integrals can be expressed uniquely in terms of these invariants, multiplying a second rank tensor.
In our examples, the tensor structures of $F_{ij}$ in (26) is  always $P_i Q_j$, $P_i Q'_j$, $P'_iQ_j$ or $P'_iQ'_j$. We do not assume that the  external gluons are transverse to their momentum $\textbf{K}$, so  we cannot replace $P'_i$ by $P_i$ or $Q'_j$ by $Q_j$. And we do find  that the $P_iQ_j$ term and the $P_iQ'_j$ terms separately give the combination (11), so it is convenient to
treat these two tensors separately (using the change of variables (27), $P'_iQ'_j$ can be exchanged for $P_iQ_j$ and similarly 
$P'_iQ_j$ can be exchanged for $P_iQ'_j$). In the examples we present, we choose only the $P_iQ_j$ terms.

There are very many graphs. In all the examples we have studied, we find that the graphs may be combined into sets such that their sum has the convergence factor $D$ in  (11).  The rest of the paper
is devoted to exhibiting some of these examples.

The existence of convergent sets when $\t \neq 0$ is a stronger condition than for $\t=0$. For example, we shall show that
graphs with denominators
\begin{equation}
\frac{1}{\P^2\Q^2\P'^2\Q'^2}
\end{equation}
and with denominators
\begin{equation}
\frac{1}{\P^2\Q^2\P'^2\Q'^2\R^2}
\end{equation}
each separately fall into convergent sets. But a term in (31) with numerator $\r^2$ is indistinguishable from (30) after the limit
$\t \rightarrow 0$ has been taken. In a previous paper \cite{aajct3}, we worked with $\t=0$; but, owing to the incorrect omission of a few graphs,
we reached a  wrong (negative) conclusion. The present paper therefore supersedes \cite{aajct3}.

In the next sections we display the examples we have studied. They are characterized by the number of their denominators, $\P^2$ etc., and also by  the quark loop they contain which is usually expressed in terms of the gluon self-energy quark loop $S(r)$ or $S(p')$.

\section{Four denominators with $S(r^2)P_iQ_j$}
\begin{figure}[t]
	\centering
		\includegraphics[width=0.75\textwidth]{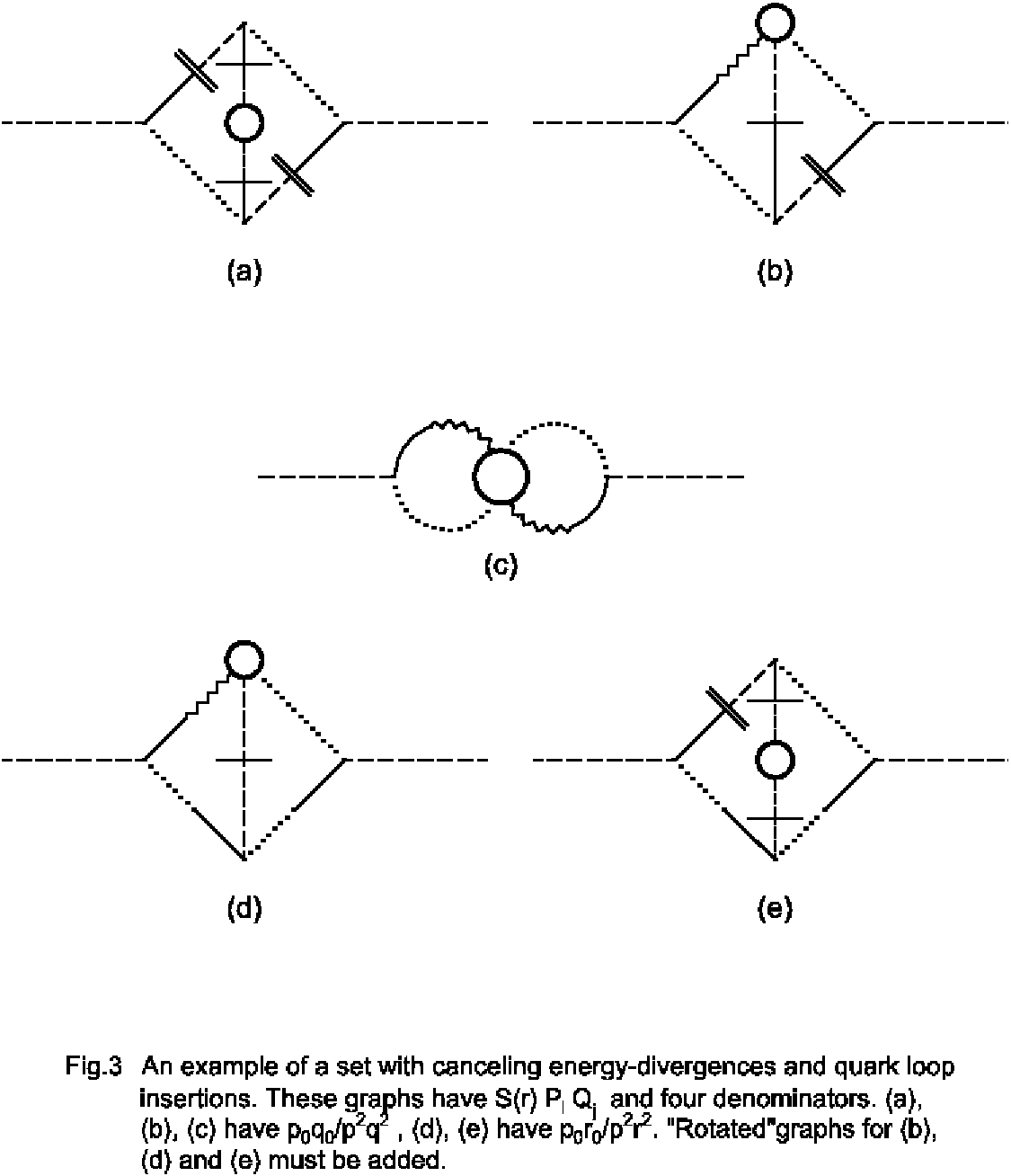}
	\label{fig:Fig3.eps}
\end{figure}

The relevant Feynman graphs are shown in Fig.3.  Our method is to use the Ward identities in the Appendix. In graphs (b) and (d) we use (A.6),  relating the 3-gluon quark loop to the self-energy.  For graph (c) we use (A.8). For example, in graph (b) there is a factor
\begin{equation}
\frac{r_0}{r^2}\left[\frac{p_0}{p^2}P_iV^{ij0}(p,r,q')+V^{0j0}(p,r,q')\right],
\end{equation}
and this is equal to
\begin{equation}
g\V[S^{j0}(r)-S^{j0}(q')]+\frac{r_0 \p^2}{r^2p^2}V^{0j0}(p,r,-p-r)
\end{equation}
where the first two terms are the result of the Ward identity (A.6), and the last term is a correction. But this last term is zero
by a similar argument to that used about equation (23) at the end of section 4: it is multiplied by a function of $\t^2 r_0^2$,
so the integrand is odd under $p_0,r_0 \rightarrow -p_0,-r_0 $.  In this section, we need only the $S^{j0}$ term in (33).
Similar arguments apply to the use of the Ward identity in (d) and (e)

The result of all this has the form
\begin{equation}
\c\int d\p d\q \int dp_0dq_0 S(r) \frac{1}{\P^2\Q^2\P'^2\Q'^2}[G_a+G_b+G_c+G_d+G_e]
\end{equation}
(corresponding to  the five graphs in Fig.3).  We must include also the 'rotated' graphs of (b), (d) and (e), that is graphs with the internal lines rotated through 180 degrees in the plane of the paper, keeping the external lines fixed. This amounts to making the substitutions
\begin{equation}
p,q,r \rightarrow -q, -p, -r.
\end{equation}

The individual contributions are
\begin{equation}
G_a=-\p.\q\U,
\end{equation}
\begin{equation}
G_b=-(\p.\r+\q.\r)\U, 
\end{equation}
\begin{equation}
G_c=-\r^2\U.
\end{equation}
\begin{equation}
G_d=\p'.\r \V + \q'.\r\W,
\end{equation}
\begin{equation}
G_e=\p.\p'\V+\q.\q'\W.
\end{equation}
Thus (34) becomes
\begin{equation}
\c\int d\p d\q \int dp_0dq_0 S(r) \frac{X_1}{\P^2\Q^2\P'^2\Q'^2}\left[\U+\V+\W \right],
\end{equation}
where
\begin{equation}
X_1 \equiv -\p'.\q' = -\frac{1}{2}[\k^2+\r^2-\p^2-\q^2],
\end{equation}
or, in  the notation (12), 
\begin{equation}
\c\int d\r d\p \int dp_0dq_0 S(r_0^2-\r^2)D\frac{X_1}{\P^2\Q^2\P'^2\Q'^2}.
\end{equation}
and, since the energy integrals are now convergent, the limit $\t \rightarrow 0$ can be taken if we wish,
giving finally
\begin{equation}
 \c \int d\r d\p \frac{X_1}{\p^2\q^2\p'^2\q'^2}\int dr_0 dp_0 S(r_0^2-\r^2)D.
\end{equation}
In the final factor in (44), the $p_0$-integral may be evaluated, giving
\begin{equation}
-i\pi \c \int d\r d\p \frac{X_1}{\p^2\q^2\p'^2\q'^2}\int dr_0 \frac{|\textbf{P}|+|\textbf{Q}|}{(r_0-k_0)^2-(|\textbf{P}|+|\textbf{Q}|)^2+i\eta }S(r_0^2-\r^2),
\end{equation}
which is a sort of $O(\hbar^3)$ correction to (22); but unlike (22) it is not independent of $k_0$, that is, in Fourier transform, it is not instantaneous. 
  
\section{Five denominators with $S(r)$}

\begin{figure}[t]
	\centering
		\includegraphics[width=0.75\textwidth]{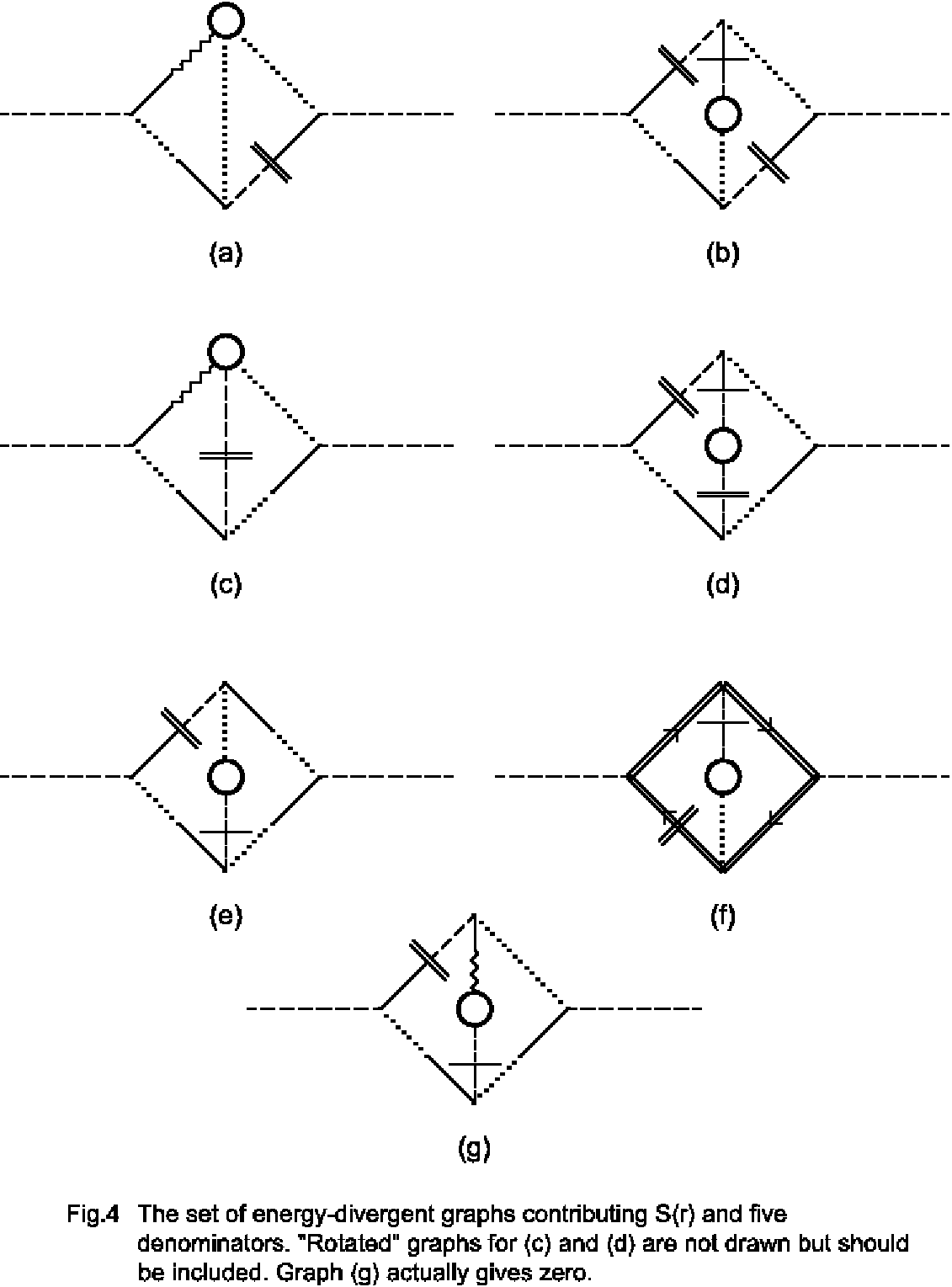}
	\label{fig:Fig4.eps}
\end{figure}

In this section, we are concerned with graphs contributing a factor
\begin{equation}
S(r)P_i Q_j /(\P^2\Q^2\P'^2\Q'^2\R^2).
\end{equation}
There are  six graphs  shown in Fig.4.  For graphs (a) and (c) we again omit convergent integrals like the last term in (33), because they are zero.
Then the graphs  Fig.4(a), (b) (and 'rotated graphs'), give
\begin{equation}
[(\q.\p' +\p.\q')\r^2]\U + [\q.\p' \p.\r +\p.\q' \q.\r]\U.
\end{equation}
This is equal to
\begin{equation}
X_2 \U \equiv (1/2)[\r^4-\r^2(\p^2+\q^2)+\p^2\q^2-\p'^2\q'^2]\U.
\end{equation}
The graphs in Fig.4 (c), (d), (e), (f), contributing respectively
\begin{equation}
[-\p'.\r \r^2 - \p'.\r \p.\r - \p'.\r \p.\q']\V + \p'^2 \q'.\r \v'.
\end{equation}
For the same reason as in (23), we may replace $\v'$ by $\V$ in (45) (since $S(r^2)$ is an even function of $r_0$ it does not affect the argument). Then the total of the terms in (49) is again proportional to $X_2$.  Finally there are contributions with $\W$, given by the substitution (35), and which therefore have the same factor $X_2$ since it is symmetric under (35).
Thus the conclusion is that the graphs in Fig.4 are a convergent set containing the convergence factor $D$ in (11) and so a convergent energy integral (10). The final form
\begin{equation}
\c\int d\p d\q \int dp_0 dr_0 S(r_0^2-\r^2)DX_2[\P^2\Q^2\P'^2\Q'^2\R^2]^{-1}
\end{equation}
using the notations of (11) and (48). Again, because of the convergence factor $D$, the limit $\t \rightarrow 0$ may be taken in (50),
giving
\begin{equation}
\c\int d\p d\q \frac{X_2}{\p^2\q^2\p'^2\q'^2\r^2}\int dr_0 dp_0 D S(r_0^2-\r^2).
\end{equation}
\def\X{|\textbf{P}|}
\def\Y{|\textbf{Q}|}
\def\Z{|\textbf{R}|}
\def\x{|\textbf{P'}|}
\def\y{|\textbf{Q'}|}
\def\z{|\textbf{R'}|}
Doing the $p_0$ integration gives
\begin{equation}
-i\pi \c\int d\p d\q \frac{X_2}{\p^2\q^2\p'^2\q'^2\r^2}\int dr_0\frac{\X+\Y}{(r_0-k_0)^2-(\X+\Y)^2+i\eta}S(r_0^2-R^2).
\end{equation}
Finally, we mention the graph in Fig.4(g). This contains a factor
\begin{equation}
\V[R_iS^{ij}(r)+r_0S^{0l}(r)]-\frac{p_0}{p^2}\frac{\r^2 r_0}{(r^2)^2}S(r).
\end{equation}
In (53), the square bracket   vanishes because of the Ward identity (A.5).
In the second term, the $r_0$-integration (for fixed $p_0$) is convergent, and so in the denominator in (46) we may set $\t=0$ except
where it is multiplied by $p_0$. The result is an odd function of $p_0$, so its integral is zero. Thus (g) contributes nothing.
In future, we omit all graphs containing  a line like the $r$-line in Fig.4(g).
\section{Six denominators with $S(r)$}
\begin{figure}[t]
\centering
	\includegraphics[width=0.75\textwidth]{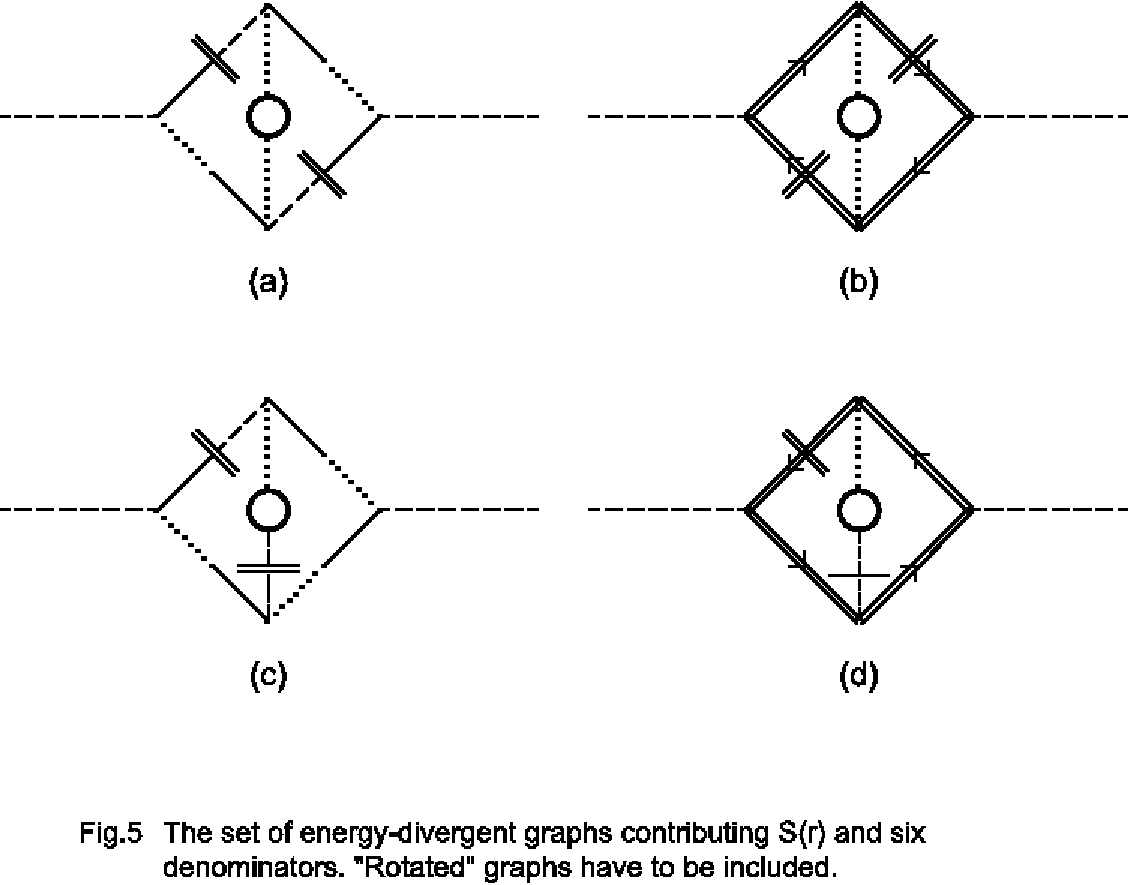}
\end{figure}
In this section we treat graphs which contribute a factor
\begin{equation}
S(r)P_i Q_j/[\P^2\Q^2\P'^2\Q'^2(\R^2)^2].
\end{equation}
There are six graphs, of which four are   shown in Fig.5 and the rotations (given by (35)) of (c) and (d) are to be included. The contributions from Fig.5 are, in order,
\begin{equation}
\r^2\left[-\p.\q' \p'.\q\U+\p'^2\q'^2\u' +\p.\q'\r.\p'\V
 - \q'.\r\p'^2\v'+\q.\p'\r.\q'\W-\p'^2\q'^2\w'\right].
\end{equation}
Neglecting at first the differences $\u'-\U$, $\v'-\V$, $\w'-\W$ (to be discussed below),
the total is then
\begin{equation}
S(r)P_i Q_j DX_3[\p^2\q^2\p'^2\q'^2(\r^2)^2]^{-1}
\end{equation}
where $D$ is defined in (11) and
\begin{equation}
X_3=\r^2[\p'^2\q'^2-\p.\q'\q.\p']$$
$$=\frac{1}{4}\r^2[-\r^4+\r^2(\p^2+\q^2+\p'^2+\q'^2)-(\p^2\q^2+\p^2\p'^2+\q^2\q'^2-3\p'^2\q'^2)].
\end{equation}
Thus the convergence factor $D$ again appears, and we can put $\t =0$ and integrate (54) over $p_0$  (for fixed $r_0$), so getting
\begin{equation}
-i\pi \c\int d\p d\q X_3[\p^2\q^2\p'^2\q'^2(\r^2)^2]^{-1}\int dr_0 \frac{\X+\Y}{(r_0-k_0)^2-(\X+\Y)^2-i\eta}S(r_0^2-\r^2),
\end{equation}
similarly to (52).

The error in neglecting $\v'-\V$ and $\w'-\W$ is zero, for the same reason as in (23) and (49). The error from $\u'-\U$ is  convergent when
$\t=0$, but is
\begin{equation}
\c\int dp dq \left[\u'-\U\right]S(r_0^2-\r^2)\int d\textbf{P} d\textbf{Q} \p'^2\q'^2\r^2[\p^2\q^2\p'^2\q'^2(\r^2)^2]^{-1}.
\end{equation}
which is finite but not zero. We  then do the $p_0$ integral (for fixed $r_0$), giving
\begin{equation}
-i\pi \c \int d\p d\q \p'^2\q'^2\r^2[\p^2\q^2\p'^2\q'^2(\r^2)^2]^{-1}$$
$$\times\int dr_0  \left[\frac{\x+\y}{(r_0+k_0)^2-(\x+\y)^2}-\frac{\X+\Y}{(r_0-k_0)^2-(\X+\Y)^2}\right]S(r_0^2-\r^2).
\end{equation}
This correction term vanishes if the 4-vector $k=0$.
Finally we combine (60) with (58), using the definition in the first line of (57) of $X_2$ to get the full result
\begin{equation}
-i\pi \c \int d\p d\q [\p^2\q^2\p'^2\q'^2(\r^2)^2]^{-1}$$
$$\times\int dr_0 \r^2 \left[\p'^2\q'^2 \frac{\x+\y}{(r_0+k_0)^2-(\x+\y)^2}-\p.\q'\q.\p'\frac{\X+\Y}{(r_0-k_0)^2-(\X+\Y)^2}\right]S(r_0^2-\r^2).
\end{equation}
This is the complete expression for the sum of the graphs in Fig.5.

\section{Four denominators with $S(p')$}
\begin{figure}[t]
\centering
	\includegraphics[width=0.75\textwidth]{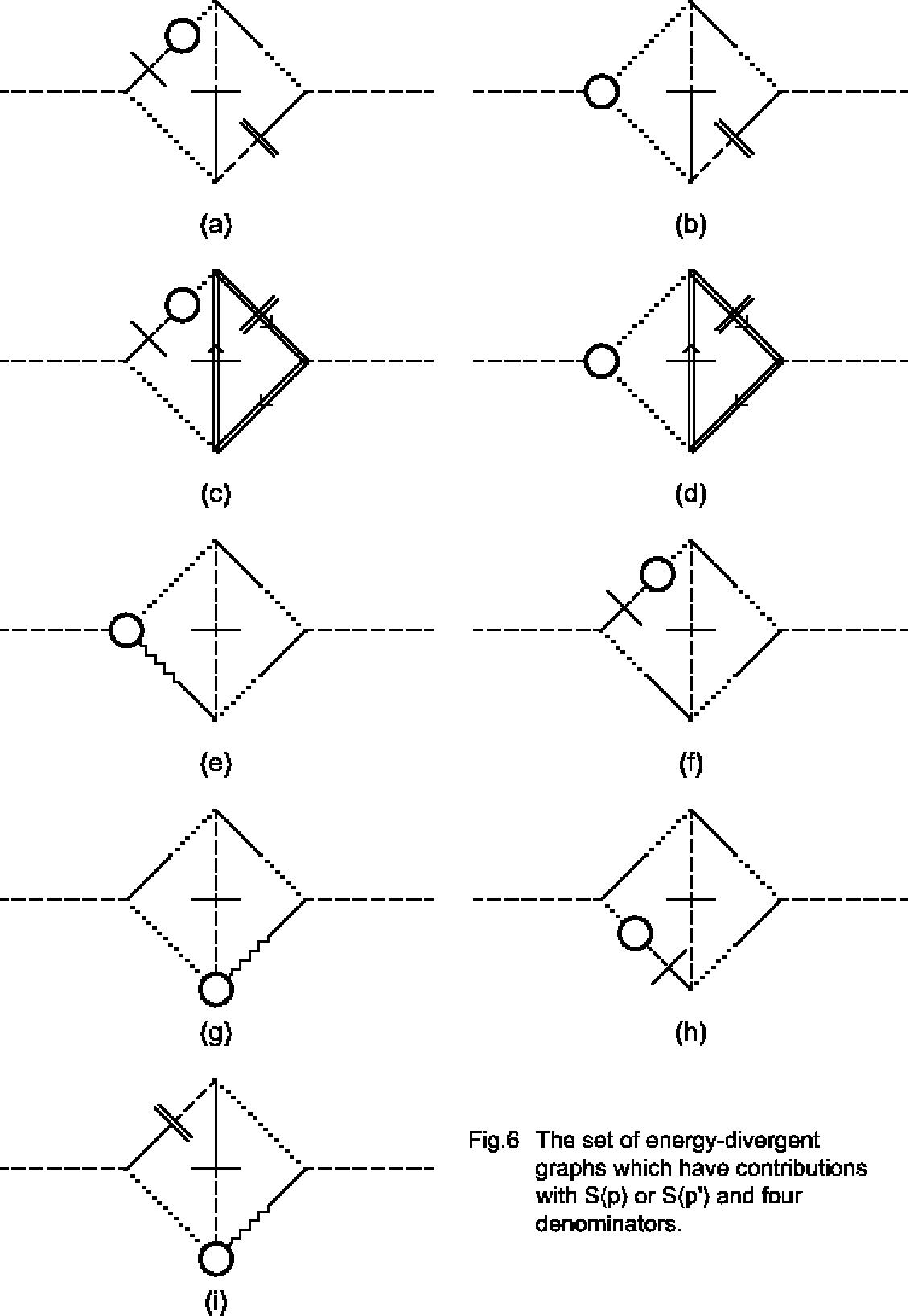}
	\label{fig:Fig6.eps}
\end{figure}
\begin{figure}[t]
	\centering
		\includegraphics[width=0.75\textwidth]{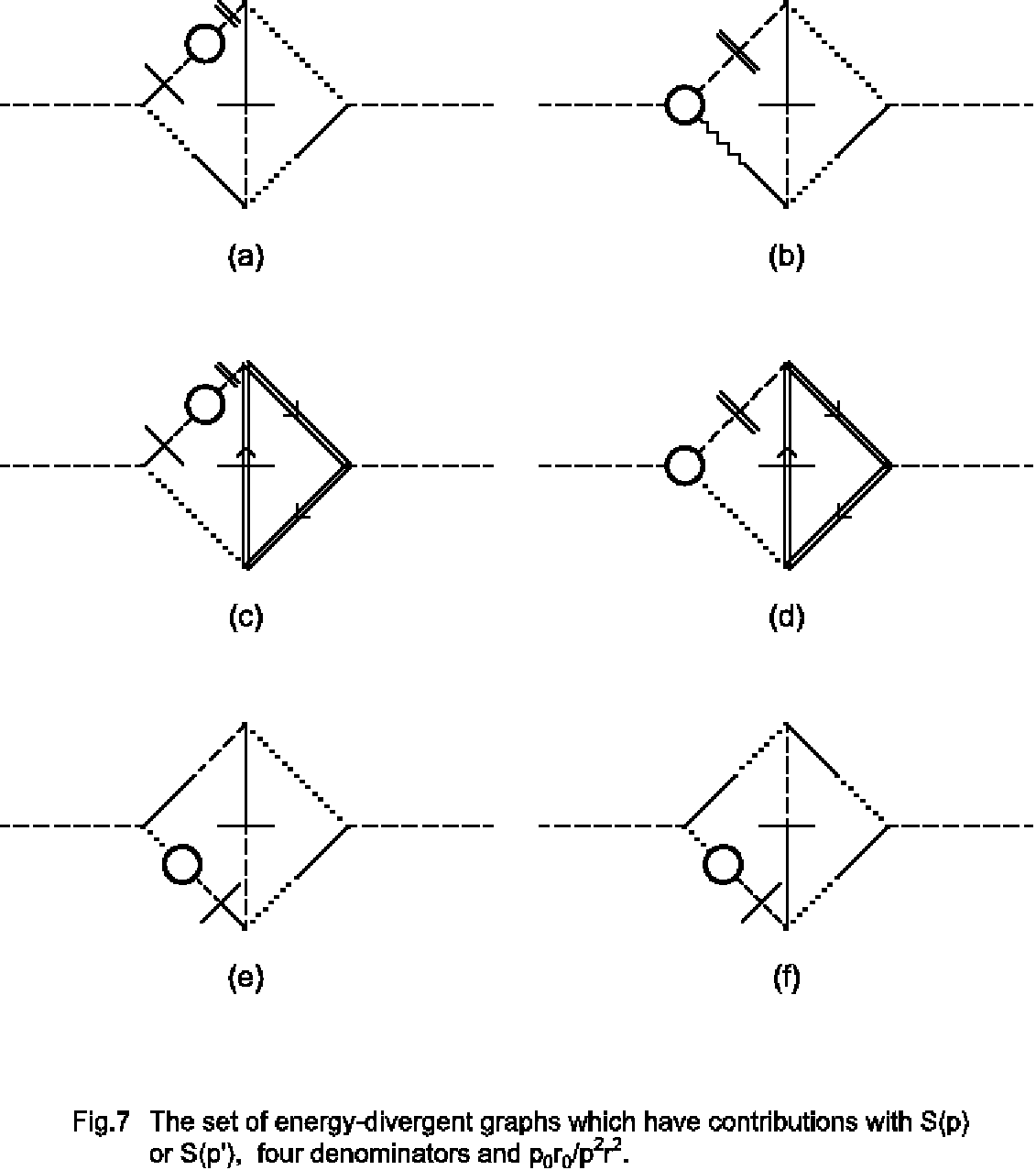}
	\label{fig:Fig7.eps}
\end{figure}

In this section we collect divergent graphs which contribute to
\begin{equation}
S(p')P_i  Q_j [\P^2\Q^2\Q'^2\P'^2]^{-1}.
\end{equation}
There are eleven logarithmically divergent graphs, shown in Figs.6 and 7. We show that when they are combined the divergences cancel. We collect part of the resulting convergent integrals, but do not attempt to find all the convergent parts (as we did in the previous sections).
For several of the graphs, the divergent parts cancel between pairs. These pairs are Fig.6 (a) and (b), (c) and (d), (e) and (f), and Fig.7
(a) and (b), (c) and (d). These cancellations stem from the Ward identity connecting quark loop vertex part $V_{\lambda\mu\nu}$
to the self energy $S_{\mu\nu}$ given in the appendix equation (A.6) and from the relation in (A.11). In general cancellations apply only at large internal energies.  The omitted corrections depend on $V$ as well as on $S$.

The remaining graphs Fig.6 (g), (h) and (i) contribute factors
\begin{equation}
-S(p')\p'.\q'\left[\frac{p'_0q_0}{q^2r^2}+\frac{1}{r^2}\right] +S(p')\p.\p'\frac{r_0p'_0p_0q_0}{r^2p^2q^2}.
\end{equation}
Neglecting convergent corrections, this reduces to
\begin{equation}
-S(p')\p'.\r\W.
\end{equation}
Graphs Fig7(e) and (f) give
\begin{equation}
-S(p')\p'.\r\V.
\end{equation}
Having included a zero contribution with $\U$ containing (13),
  we  get the convergent combination (having put $\t=0$)
\begin{equation}
\c\int d\p d\q [\p^2\q^2\p'^2\q'^2]^{-1}X_4 \int dp'_0 dr_0 DS(p')
\end{equation}
where $D$ is defined in (11) and
\begin{equation}
X_4=-\p'.\r= -(\q^2-\r^2-\p'^2)/2.
\end{equation}
Expression (66) gives
\begin{equation}
-i\pi \c\int d\p d\q [\p^2\q^2\p'^2\q'^2]^{-1}X_4 \int dp'_0 \frac{\Y+\Z}{{p'}_0^2-(\Y+\Z)^2}S({p'}_0^2-\p'^2)
\end{equation}
Equation (68) is independent of $k_0$ and looks like  an $O(\hbar^3)$ correction to (22): but it is not the complete contribution from Figs6 and 7, which probably involves the vertex function $V_{\lambda\mu\nu}$ as well as $S$.


\section{Five denominators with $S(p')$}
\begin{figure}[t]
	\includegraphics[width=0.75\textwidth]{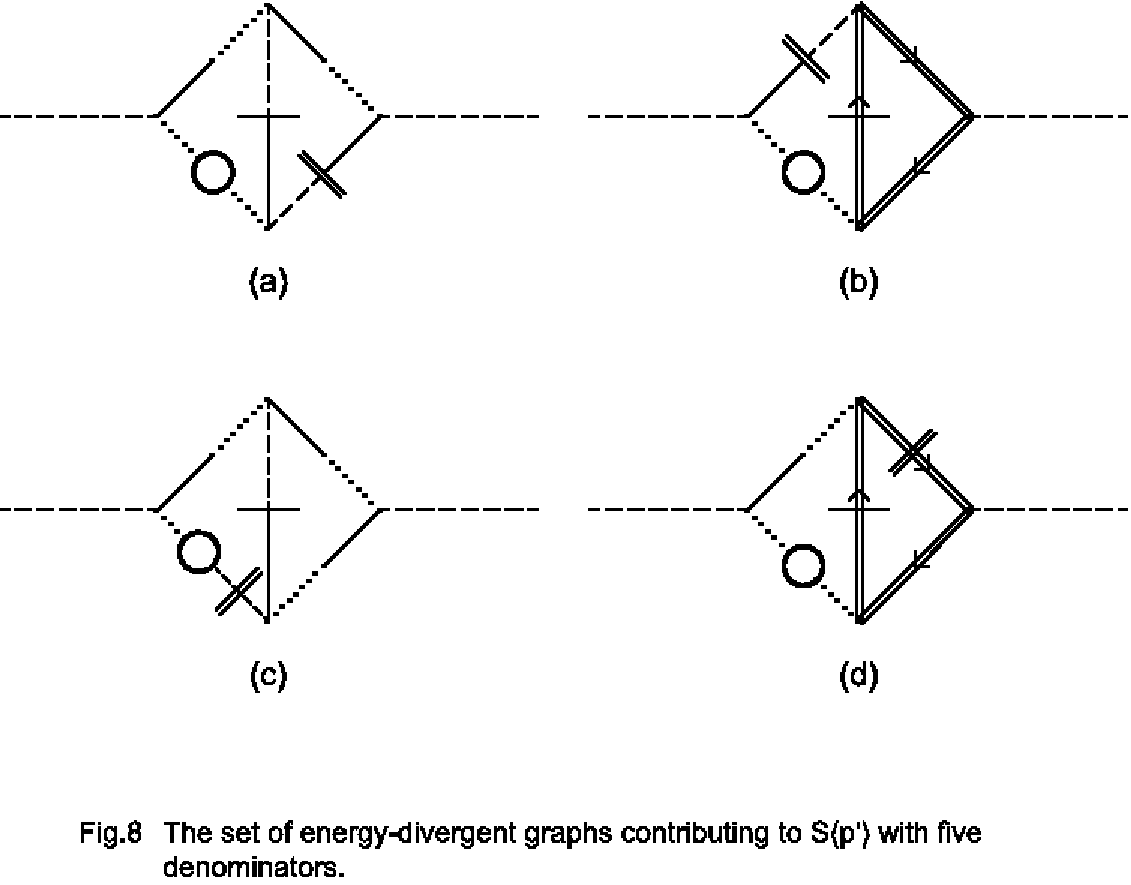}
	\label{fig:Fig8.eps}
\end{figure}

This section concerns energy-divergent graphs which contribute to
\begin{equation}
[\P^2\Q^2\Q'^2(\P'^2)^2]^{-1}S(p')P_iQ_j.
\end{equation}
There are four graphs shown in Fig.8. They contribute, in order (a), (b), (c), (d),
\begin{equation}
\p'^2\left[\q.\q'\W+\p.\q'\V-\p'.\q'\v'-\q'^2\w'\right].
\end{equation}
Neglecting the convergent contributions from $\v'-\V$ and $\w'-\W$ (discussed below), and using adding a zero term with $\U$ by using (12), we get the convergent combination
\begin{equation}
\c\int d\p d\q X_5[\p^2\q^2\q'^2(\p'^2)^2]^{-1}\int dp'_0 dr_0 D S(p'^2),
\end{equation}
where we have put $\t=0$, and
\begin{equation}
X_5 \equiv \p'^2\k.\q'=-\frac{1}{2}\p'^2(\k^2-\q^2+\q'^2).
\end{equation}
In (72), the $r_0$-integration gives
\begin{equation}
-i\pi \c\int d\p d\q X_5[\p^2\q^2\q'^2(\p'^2)^2]^{-1}\int dp'_0\frac{\Z+\Y}{{p'_0}^2-(\Z+\Y)^2}S({p'_0}^2-\p'^2)
\end{equation}
which is independent of $k_0$.

The neglected convergent correction from $\v'-\V$ is zero from the argument used about (23) (the extra factor $S(p')$ does not affect the argument).
But the convergent correction  due to $\w'-\W$ involves the integral, for fixed $p'_0$
\begin{equation}
J(p'_0,k_0)\equiv\int dr_0 [\w'-\W]=-i\pi \left[\frac{\x+\Z}{(p'_0+k_0)^2-(\x+\Z)^2}-\frac{\X+\Z}{{p'_0}^2-(\X+\Z)^2}\right],
\end{equation}
which leads to the extra factor
\begin{equation}
i\pi \c\int d\p d\q \p'^2\p'.\q' [\p^2\q^2(\p'^2)^2\q'^2]^{-1}\int dp'_0 J(p'_0,k_0) S({p'_0}^2-\p'^2)
\end{equation}
The total of (73) and (75) is
\begin{equation}
i\pi \c\int d\p d\q \p'^2 [\p^2\q^2(\p'^2)^2\q'^2]^{-1}\int dp'_0 S({p'_0}^2-\p'^2)$$
$$\times \left[ \q.\q'\frac{\Y+\Z}{{p'}_0^2-(\Y+\Z)^2}- \q'^2\frac{\y+\Z}{(p'_0+k_0)^2-(\y+\Z)^2}\right].
\end{equation}

The factors $\p'^2$ in numerator and denominator cancel, but we leave them there as a reminder of their origins.

\section{Six denominators with S(p')}
\begin{figure}[t]
	\centering
		\includegraphics[width=0.75\textwidth]{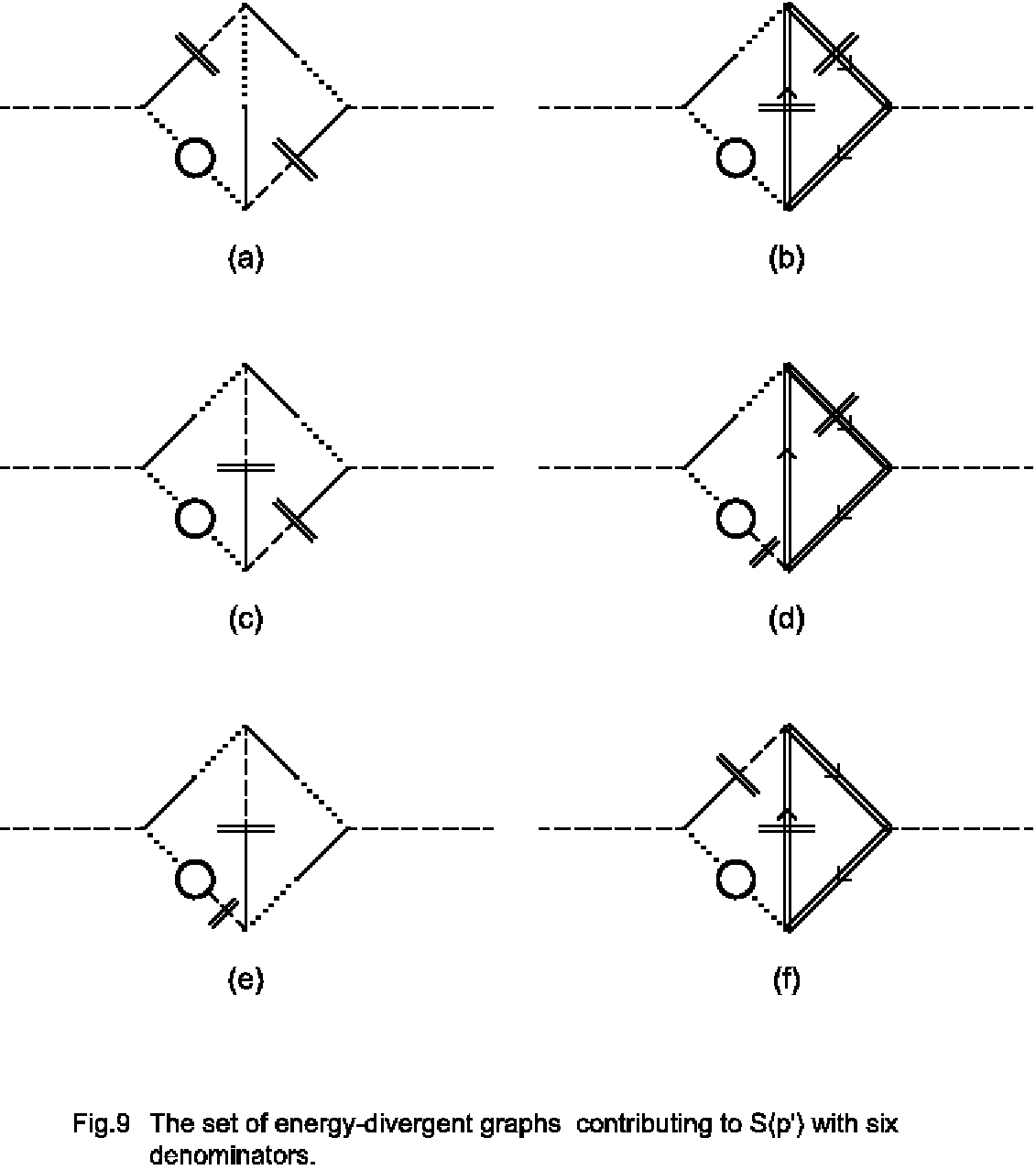}
	\label{fig:Fig9.eps}
\end{figure}

This section concerns graphs which contribute to
\begin{equation}
S(p')[\p^2\q^2(\p'^2)^2\q'^2\r^2]^{-1}P_iQ_j.
\end{equation}
These graphs are shown in Fig.9. The factors are, in order,
\begin{equation}
\p'^2\left[\p.\q' \q.\r \U+\q'^2\r^2\w' -\q.\r\q'.\r\W-\q'^2\p'.\r \u' +\p'.\r \q'.\r\v'-\p.\q'\r^2\V\right].
\end{equation}
Neglecting the differences $\u'-\U$, $\v'-\V$ and $\w'-\W$, the above expression gives
\begin{equation}
X_6 D\equiv \p'^2(\r^2\q'^2-\r.\q\r.\q')D=\frac{1}{4}\p'^2[-\r^4+\r^2(\p^2+\p'^2 -\q^2+3\q'^2)-(\p^2-\q'^2)(\q^2-\p'^2)]D.
\end{equation}
Because of the convergent combination $D$ (defined in (11)), we can set $\t=0$ and get the result
\begin{equation}
\gamma \int d\p d\q X_6[\p^2\q^2(\p'^2)^2\q'^2\r^2]^{-1}P_i Q_j\int  dp'_0 dr_0  D S({p'_0}^2-\p'^2)
\end{equation}
\begin{equation}
=-i\pi \gamma \int d\p d\q X_6[\p^2\q^2(\p'^2)^2\q'^2\r^2]^{-1}P_i Q_j\int dp'_0 \frac{\Y+\Z}{{p'_0}^2-(\Y+\Z)^2}S({p'_0}^2-\p'^2),
\end{equation}
which is independent of $k_0$.

We now study the corrections due to the neglected convergent terms.  That due to neglect of  $\v'-\V$ is zero by the same argument as used for (23), with interchange of the roles of $r$ and $p'$. Also, we can write
\begin{equation}
\u'-\U= \frac{p'_0}{p'^2}\left[\frac{q'_0}{q'^2}-\frac{q_0}{q^q}\right]+\frac{q_0}{q^2}\left[\frac{p'_0}{p'^2}-\frac{p_0}{p^2}\right].
\end{equation}
For the first term in (82), the $q_0$-integration is convergent, so we can put $\t=0$ except in $\t p'_0$, and the integrand is then an odd function of $p'_0$. For the last term in (82), the $p'_0$-integration is convergent, so we can put $\t=0$ except in $\t q_0$,
and the integrand is an odd function of $q_0$. Thus the correction terms in (82) are zero.
The correction due to $\w'-\W$ contains the  factor
\begin{equation}
-i\pi\q'^2\r^2\int dp'_0 \left[ \frac{\y+\Z}{(p'_0+k_0)^2-(\y+\Z)^2}-\frac{\Y+\Z}{{p'_0}^2-(\Y+\Z)^2}\right] S({p'_0}^2-P'^2).
\end{equation}
Combining this with (81), the complete result is
\begin{equation}
-i\pi \gamma \int d\p d\q [\p^2\q^2(\p'^2)^2\q'^2\r^2]^{-1}\p'^2P_i Q_j$$
$$\times \int dp'_0 \left[ \q'^2\r^2  \frac{\y+\Z}{(p'_0+k_0)^2-(\y+\Z)^2}-\q.\r \q'.\r \frac{\Y+\Z}{{p'_0}^2-(\Y+\Z)^2}\right]S({p'_0}^2-\p'^2).
\end{equation}
Again we purposely refrain from canceling $\p'^2$ factors.

\section{Conclusion}
In Coulomb gauge QCD perturbation theory, at three loop order,  there are energy-divergences from individual Feynman graphs.
We have given six examples showing that these divergences cancel out when the relevant graphs are combined. We use an interpolating gauge
to control the divergences at intermediate stages of the calculations.
(We have actually checked some more examples which we have not put in this paper.) So we conjecture that this is  a general property.
 These divergence cancellations  are not implied  by gauge-fixing independence (in our case, independence of the parameter $\t$ in (1)), because the amplitudes we study are not $S$-matrix elements and so are not expected to be gauge-independent.

We have not constructed a general proof, but there are properties which would play a part in making such a proof.
For example, in Figs.(6) and (7), the cancellations between the divergences in seven pairs of graphs are  simple consequences of the Ward identities, as
expressed in (A.6), (A.11) and (A.12). Moreover, similar  cancellations would occur in graphs where an arbitrary number of external gluons was attached
to the left sides of Fig.6(a), (b), (c) and (d) (for example), keeping the top and bottom vertices unchanged. The left side
would then constitute a complete \emph{chain} in the sense used in \cite{aajct2} (in the present example, the chain would be the equal ghost propagator).

In five of our examples, we derive a compact expression for the sum of the set of graphs, in terms of a single scalar function, $S$,
which is the gluon self-energy from a quark loop. These expressions look like radiative corrections to term which  Christ and Lee \cite{christlee} argue should be added to the Coulomb gauge Hamiltonian, except that they are not in general instantaneous (that is energy-independent). Since we have no other way to calculate these $O(\hbar^3)$ terms, there seems no alternative but to combine sets of graphs as we have done here.
It would be interesting to know whether these complications are reflected in non-perturbative QCD, for instance in lattice calculations in the Coulomb gauge.

The sets of graphs which have to be combined have similar integrands in the  interpolating gauge, but, after we take the limit ($\t =0$)
to the Coulomb gauge, contributions from different sets may give the same integrand, as illustrated by the example in (30) and (31).
So terms containing
\begin{equation}
S(r)P_iQ_j[\p^2\q^2\p'^2\q'^2]^{-1}
\end{equation}
get contributions from $X_1$ in (42), from $X_2$ in (48), and from $X_3$ in (57) as follows:
\begin{equation}
[-\k^2/2 - \r^2/2]+[\r^2/2] +[-\r^2/4].
\end{equation}
(The correction term (60) does not contribute to (85).)
Thus the final result for terms contributing to (85) is
\begin{equation}
i\pi \c\int d\p d\q[(\k^2/2)+(\r^2/4)] P_iQ_j[\p^2\q^2\p'^2\q'^2]^{-1}\int dr_0\left[\frac{\X+\Y}{(r_0-k_0)^2-(\X+\Y)^2}S(r_0^2-\r^2) \right],
\end{equation}
which may be considered as a higher-order correction to (22) (which is not however independent of $k_0$).
 \section{Appendix: quark loops and Ward identities}
We review the notation we used in\cite{aajct3}. The gluon self-energy function from a quark loop is
$$
(2\pi)^4S_{\mu\nu}(p)=(2\pi)^4C_q(p_{\mu}p_{\nu}-p^2g_{\mu\nu})S(p^2) \eqno(A.1)$$
where
$$
S=g^28i\pi^{d/2}\Gamma(\epsilon)\frac{[\Gamma(d/2)]^2}{\Gamma(d)} [(-p^2-i\eta)^{-\epsilon}-(\mu^2)^{-\epsilon}]
\eqno(A.2)
$$
where $d=4-2\epsilon$ is the space-time dimension and $\mu$ is the arbitrary unit of mass. The vertex part from a quark triangle is
$$
\texttt{tr}(t^a[t^b , t^c])V_{\lambda\mu\nu}(p,q,r)(2\pi)^4 \delta^4(p+q+r)
\eqno(A.3)$$
and the four gluon amplitude from quark squares is
$$
\texttt{tr}(t^a t^b t^c t^d+t^d t^c t^b t^a)W_{\lambda\mu\nu\sigma}(p,q,r,s)(2\pi)^4\delta^4(p+q+r+s),
\eqno(A.4)$$
$t^a$ etc being colour matrices in the fundamental representation and $C_q$ the Casimir of that representation. The momenta are defined to be IN to the vertices.

The Ward identities are:
$$
p^{\mu}S_{\mu\nu}(p)=0,
\eqno(A.5)$$
$$
p^{\lambda}V_{\lambda\mu\nu}(p,q,r)=g[S_{\mu\nu}(q)-S_{\mu\nu}(r)],
\eqno(A.6)$$
$$
p^{\lambda}W_{\lambda\mu\nu\sigma}(p,q,r,s)=g[-V_{\mu\nu\sigma}(p+q,r,s)+V_{\mu\nu\sigma}(q,r,p+s)].
\eqno(A.7)$$
Combining (A7) with (A6), we get
$$
p^{\mu}q^{\nu}W_{\mu 0 \nu 0}(p,q',-q,-p')=g^2[\p'^2S(p')+\q'^2S(q')-\r^2S(r)-\k^2S(k)].
\eqno(A.8)$$

In the high-energy approximation (which is all that is necessary to prove convergence), we may use
$$
p^{\mu}S_{\mu i}(p)\approx p_0S_{0i},
\eqno(A.9)$$
$$
p^{\lambda}V_{\lambda i j}(p,q,r) \approx p_0V_{0ij}(p,q,r).
\eqno(A.10)$$

Using the Ward identities, we have \cite{aajct3} found high-energy approximations for some of  the $V$ and $W$ amplitudes in terms of $S$.
For example
$$
V_{00i}(p,-p',k) \approx g[P'_i S(p')+P_i S(p)],
\eqno(A.11)$$
and
$$
V_{0ij}(p,-p',k) \approx g\delta_{ij}p_0S(p) \eqno(A.12)
$$
under the additional conditions
$$
|p_0|, |p'_0|>> k_0,
\eqno(A.13)$$
The above are the results which we use, especially in Section IX.\\

\noindent{Acknowledgement.
We are grateful to Dr. D. Sokcevic for making the Latex version perfect.}

\end{document}